\documentclass{article}
\usepackage{graphicx}
\usepackage{amsmath}

\input{tcilatex}
\begin{document}

\title{The Unitary Transformation in Quantum Teleportation}
\author{Zheng-Chuan Wang \\
Department of Physics, The Graduate School of the Chinese \\
Academy of Sciences, P. O. Box 4588, Beijing 100049, China.}
\maketitle

\begin{abstract}
In the well known treatment of quantum teleportation, the receiver should
convert the state of his EPR particle into the replica of the unknown
quantum state by one of four possible unitary transformations. However, the
importance of these unitary transformations must be emphasized. We will show
in this paper that the receiver can not transform the state of his particle
into an exact replica of the unknown state which the sender want to transfer
if he have not a proper implementation of these unitary transformations. In
the procedure of converting state, the inevitable coupling between EPR
particle and environment which is needed by the implementation of unitary
transformations will reduce the accuracy of the replica.

03.67.Hk, 03.67.-a, 03.65.Ta..
\end{abstract}

In 1993, Bennett et al.\cite{bennett} proposed a famous treatment to
transfer an intact quantum state from one place to another by use of the
long-range correlation between EPR pair of particles. In their scheme, an
unknown state and one of EPR particles are given to the sender, and the
sender then perform a complete measurement on the joint system of the
unknown state and her EPR state. After this, the receiver will perform a
unitary transformation on the second particle of the EPR pair to obtain the
replica of the unknown state, certainly, this unitary transformation is
determined by the results of measurement told by the sender through a
classical channel. The experimental realizations of their treatment were
exhibited by Bouwmeester et al.\cite{bouw} and Boschi et al.\cite{boschi},
respectively, in which an initial photon which carries the polarization is
transferred by use of a pair of entangled photons prepared in an EPR state.
These approaches to quantum teleportation had inspired many investigations
into this field, such as the discussions of continuous variable quantum
teleportation\cite{furu,bra,yon}, the analysis of quantum fluctuation in the
teleportation\cite{pol} etc.

However, it should be pointed out that the physical implementation of the
unitary transformations on the second EPR particle should be noticed, they
are not mere the pure mathematical transformations, we need realize these
unitary transformations by other physical systems. For convenience, we
summarily call these system $^{\prime }$environment$^{\prime }$. The
inevitable interaction between EPR particle in the hands of receiver and the
environment will affect the replica of the unknown state. Even worse, if we
have not properly chosen the physical realization of these unitary
transformations, the unknown quantum state will not be teleported enough
accurately because of the influence of environment.

Suppose a sender, traditionally called $^{\prime }$Alice$^{\prime }$, who
wish to communicate an unknown quantum state $|\Psi \rangle =a|0\rangle
+b|1\rangle $ of spin-1/2 particle (particle 1) to a receiver, $^{\prime }$%
Bob$^{\prime }$. Two other spin-1/2 particles are prepared in an EPR singlet
state. According to Bennett et al.$^{\prime }$s treatment, one EPR particle
(particle 2) is given to Alice, the other (particle 3) is given to Bob.
Alice makes a combined measurement on her EPR particle 2 and the unknown
particle 1, then Bob$^{\prime }$s particle 3 will be in one of the following
four pure states: -$a|0\rangle _{3}-b|1\rangle _{3}$, -$a|0\rangle
_{3}+b|1\rangle _{3}$, $b|0\rangle _{3}+a|1\rangle _{3}$, and -$b|0\rangle
_{3}+a|1\rangle _{3}$. In the ideal case, Bob can convert the state of
particle 3 into an exact replica of the initial state $|\Psi \rangle
=a|0\rangle +b|1\rangle $ by a unitary transformation which depends on the
results of measurement told by Alice via classical channel. However, these
unitary transformations must be performed through other physical systems or
apparatus, which can be summarily described as $^{\prime }$environment$%
^{\prime }$, then the interaction between particle 3 and the $^{\prime }$%
environment$^{\prime }$ occurs, which will couple the quantum state of
environment with particle 3 and violate the accurate replica of the initial
quantum state $|\Psi \rangle $.

In fact, the above interaction between particle 3 and environment makes the
quantum state of the combined system (particle 3-environment) evolves as
follows\cite{zurek}

\begin{eqnarray}
|\Phi (t &=&t_{0})\rangle =|E_{0}\rangle \otimes |\Psi \rangle _{3} \\
&\longrightarrow &|\Phi (t>t_{1})\rangle =C_{0}a|E_{0}\rangle |0\rangle
_{3}+C_{1}b|E_{1}\rangle |1\rangle _{3}.  \notag
\end{eqnarray}
In the above, $|E_{0}\rangle $, $|E_{1}\rangle $ are the state vectors of
environment, while $|\Psi \rangle _{3}$ describes the quantum state of
particle 3. As a result of particle 3-environment interaction, the
correlation between particle and environment has been established after time 
$t_{1}$, the state vectors of particle 3 and environment have coupled to
each other after time $t_{1}$. Expression (1) clearly demonstrates the
violation of pure state $|\Psi \rangle _{3}$ after the practical
implementation of unitary transformations. We can also show this violation
by its density matrix. The reduced density matrix of particle 3 is

\begin{eqnarray}
\rho _{3} &=&Tr_{E}[|\Phi (t>t_{1})\rangle \langle \Phi (t>t_{1})|] \\
&=&(|C_{0}a|^{2}+|C_{0}a|^{2}|\langle E_{1}|E_{0}\rangle |^{2})|0\rangle
\langle 0|+(2C_{0}C_{1}^{\ast }ab^{\ast }\langle E_{1}|E_{0}\rangle
)|0\rangle \langle 1|  \notag \\
&&+(2C_{1}C_{0}^{\ast }ba^{\ast }\langle E_{0}|E_{1}\rangle )|1\rangle
\langle 0|+(|C_{1}b|^{2}+|C_{1}b|^{2}|\langle E_{0}|E_{1}\rangle
|^{2})|1\rangle \langle 1|.  \notag
\end{eqnarray}
When the state vectors $|E_{0}\rangle $, $|E_{1}\rangle $ of environment are
orthogonal to each other, the density matrix can reduce to

\begin{equation}
\rho _{3}=|C_{0}a|^{2}|0\rangle \langle 0|+|C_{1}b|^{2}|1\rangle \langle 1|,
\end{equation}
which indicates pure state $|\Psi \rangle _{3}$ has become to a mixture
state. Generally, the pure state $|\Psi \rangle _{3}$ of Bobs EPR particle
will reduce to a mixture state after a practical realization of unitary
transformations, in the end Bob can not obtain a pure state of particle 3,
and can not convert $|\Psi \rangle _{3}$ into the initial pure state $|\Psi
\rangle $ which Alice sought to teleport. An unknown quantum state can thus
not be teleported enough accurately because of the physical implementation
of unitary transformations.

In the general, there exists deviation between the replica of unknown state
in the hands of Bob after practical unitary transformation and the initial
quantum state $|\Psi \rangle _{1}$ prepared by Alice. We can evaluate the
above deviation by the difference between $\rho _{3}$ and the density matrix 
$\rho _{1}$ of pure state $|\Psi \rangle _{1}$, it is

\begin{equation}
\delta =\sqrt{\sum_{n,m}|(\rho _{3})_{nm}-(\rho _{1})_{nm}|^{2}}.
\end{equation}
Considering the density matrix $\rho _{3}$ in the above expression (2), this
deviation can be further written as

\begin{eqnarray}
\delta ^{2} &=&||C_{0}a|^{2}+|C_{0}a|^{2}|\langle E_{0}|E_{1}\rangle
|^{2}-|a|^{2}|^{2}+|2C_{0}C_{1}^{\ast }ab^{\ast }\langle E_{1}|E_{0}\rangle
-ab^{\ast }|^{2} \\
&&+|2C_{1}C_{0}^{\ast }ba^{\ast }\langle E_{0}|E_{1}\rangle -ba^{\ast
}|^{2}+||C_{1}b|^{2}+|C_{1}b|^{2}|\langle E_{1}|E_{0}\rangle
|^{2}-|b|^{2}|^{2}.  \notag
\end{eqnarray}
We can see that the deviation is determined by the state vectors of
environment. If we properly arrange the environment, hence the unitary
transformation, and choose the state vector $|E_{0}\rangle =|E_{1}\rangle $, 
$C_{0}=C_{1}=\frac{1}{\sqrt{2}}$, then the deviation will vanish. In this
special case, the state vector of combined particle 3-environment system is $%
|\Phi (t>t_{1})\rangle =\frac{1}{\sqrt{2}}|E_{1}\rangle \otimes (a|0\rangle
_{3}+b|1\rangle _{3})$, there is no correlation between state vectors of
particle 3 and environment at all, the quantum state of particle 3 still
remain in a pure state after this special realization of unitary
transformations. Certainly, that is a very special case and difficult to
realize in an experiment. Generally speaking, we can not transfer an unknown
state from one place to another enough accurately unless we choose the
special physical implementation of unitary transformations.

Superficially, Bennett et al.$^{\prime }$s treatment of quantum
teleportation have not broken the quantum non-cloning theorem\cite{wotters},
because the initial unknown state has been destroyed after this
teleportation. However, if we analyze this procedure carefully, we will find
it surely break the quantum non-cloning theorem. The key point still lie in
the physical realization of unitary transformations. In essence, quantum
non-cloning theorem is similar to our above analysis. To clone an unknown
quantum state $|\Psi \rangle =a|0\rangle +b|1\rangle $ of a spin-1/2
particle, we need the necessary apparatus and environment which are also
summarily described as general environment. The outcome of the cloning
procedure is $a|A\rangle _{0}|0\rangle |0\rangle +b|A\rangle _{1}|1\rangle
|1\rangle $, where $|A\rangle _{i}(i=0,1)$ are the quantum state vectors of
apparatus, and the new state vectors $|0\rangle $ and $|1\rangle $ in the
outcome are provided by the environment. There are correlation between the
unknown state and the state vectors of environment. The general state
vectors of environment $|A\rangle _{i}|i\rangle (i=0,1)$ just correspond to
the state vectors $C_{i}|E_{i}\rangle (i=0,1)$ of environment in expression
(1). So our above analysis is equivalent to quantum non-cloning theorem, it
is the environment that leads to the impossibility for an unknown state
being cloned accurately, the interaction of particle and environment will
cause the decoherence of pure state of particle to a mixture state. To
teleport an unknown quantum state enough accurately will eventually break
the quantum non-cloning theorem because of the influence of environment.

In summary, we have shown that an unknown quantum state can not be
teleported enough accurately from one place to another when we consider the
practical physical implementation of unitary transformations except some
special cases. The coupling of environment and particle will reduce the
accuracy of converting procedure, it is another manifest of quantum
non-cloning theorem.

\bigskip $\mathbf{Acknowledgments}$

This work is supported by the NNSF (Grant No. 10404037).

\end{document}